\documentstyle[aps,amssymb,psfig]{revtex}
\begin{document}

\title{
\rightline{\small{\em To appear in Physical Review D\/}}
Spacetime perspective of Schwarzschild lensing}

\author{Simonetta Frittelli$^{a,b}$, 
	Thomas P. Kling$^{b}$ and Ezra T.  Newman$^{b}$ \\
$^{a}$Department of Physics, Duquesne University, 
Pittsburgh, PA 15282\\ 
$^{b}$Department of Physics and Astronomy, 
University of Pittsburgh, 
Pittsburgh, PA 15260}

\date{November 29, 1999} 

\maketitle

\begin{abstract} 

We propose a definition of an exact lens equation {\it{without
reference to a background spacetime}}, and construct the exact
lens equation explicitly in the case of Schwarzschild
spacetime.  For the Schwarzschild case, we give exact
expressions for the angular-diameter distance to the sources
as well as for the magnification factor and time of arrival of
the images.  We compare the exact lens equation with the
standard lens equation, derived under the thin-lens-weak-field
assumption (where the light rays are geodesics of the
background with sharp bending in the lens plane, and the
gravitational field is weak), and verify the fact that the
standard weak-field thin-lens equation is inadequate at small
impact parameter.  We show that the second-order correction to
the weak-field thin-lens equation is inaccurate as well. 
Finally, we compare the exact lens equation with the recently
proposed strong-field thin-lens equation, obtained under the
assumption of straight paths but without the small angle
approximation, i.e., with allowed large bending angles.  We
show that the strong-field thin-lens equation is remarkably
accurate, even for lightrays that take several turns around
the lens before reaching the observer.  

\end{abstract}

\section{Introduction}

The phenomenon of gravitational lensing is firmly associated
with the physics of a four-dimensional Lorentzian spacetime
that satisfies the Einstein equations.  Yet, it has become a
common practice in the study of lensing to break with the
basic ideas of general relativity by using the linearized
Einstein equations off a fixed background, the thin lens
approximation, and treating the bending of light as a linear
phenomenon -- without mention of its connection with the full
theory.

This point of view is very much justified by the accuracy in
the comparison of contemporary observations with the resulting
calculations, i.e., general relativity does play an essential
role in lensing but the weak field approach appears to be
quite adequate for most discussions~\cite{Ehlers}.

However, it is now a fact that the strong field
characteristics of general relativity \textit{per se\/} are
observed in nature as well.  Black holes are possibly
ubiquitous~\cite{Blanford}, and a super-massive black hole may
exist in the center of every spiral galaxy.  Here is where the
full theory of general relativity takes the leading part. In
order to describe bending of light by black holes or in high
curvature regions, it is necessary to write lens equations
that respect the intrinsic nature of general relativity,
namely:  covariance and non-linearity.

The difficulty in writing down a lens equation that respects
covariance and non-linearity is very much of a conceptual
type.  In fact, even when such a lens equation is developed,
it is hard to interpret.  A spacetime containing a lens
\textit{is not} the superposition of two spaces, a background
spacetime and a lens space.  Two different spacetimes are two
different entities, and there are an infinite number of ways
of identifying them point-wise.  What is the meaning of the
angular location of a source \textit{in the absence of a
lens\/ - } an idea used extensively in the thin-lens
approximation?  What are the preferred angular coordinates
that give the thin-lens equation its meaning?  How do we refer
to the distances between the observer, the source and the lens
in a coordinate independent manner, or what is the preferred
coordinate distance to use?  All these questions have
perfectly good answers if a background spacetime is available
to us and we are given leave to isolate the lensing action
from the background.  This is not so if there is no
background.  Without reference to a background, some of these
questions have no answers, and some do not even make sense. 
Treating lensing phenomena strictly in the context of the full
theory of relativity requires other ideas and approaches.

We have recently introduced a proposal for a lens equation
without reference to a background~\cite{FN,EFN}.  An exact
lens equation on an arbitrary Lorentzian spacetime can be
written down, at least in principle, since it amounts,
basically, to finding all the light-rays that reach the eye of
an observer.  However, for it to be meaningful, it is
necessary to express the equation in such a way that it can be
used in an astrophysical context; it must be written or
expressed in terms of observable quantities.  To some extent,
we believe that we have partially succeeded in doing that.

As an illustration, we develop and interpret in full detail
our lens equation in the case of a Schwarzschild black hole;
explicitly working out quantities of astrophysical interest
for lensing, such as the angular diameter distance and
magnification factors.  Furthermore, we use our exact lens
equation to test the effectiveness of other lens equations
that can be written down in the case that a background is
available, most notably the lens equation obtained recently by
K.  S.  Virbhadra and G.  F.  R.  Ellis~\cite {ellis}.

In Sec.II we discuss the idealized situation where, in
principle, the null geodesic equations can be solved exactly
for a static metric and stationary source and show, again in
principle, how a set of lens equations can be constructed,
while in Sec.III these ideas are then applied to the
Schwarzschild black hole lensing problem.  In this section,
the important physical quantities such as the angular-diameter
distance to a source and the magnification factor are
explicitly calculated.  In the subsequent sections, we compare
the exact results with the thin lens calculations.

\section{The exact lens equation}


We begin with a four-dimensional static spacetime
$({\frak{M}},g_{ab}(x^{a}))$ with local coordinates $x^{a}$
and consider an observer, at rest in the local coordinates, on
a world-line given parametrically by $x_{0}^{a}(\tau )$,
$\tau$ being the observer's proper-time.  The observer,
looking out, sees null geodesics reaching him from all past
null directions, $l^{a}$.  These observed directions, labeled
by the spatial projections (orthogonal to the observer's
velocity vector, $v^{a}=\frac{d}{d\tau} x_{0}^{a}(\tau )$) of
the null vectors, can be taken as the two angular coordinates
of the observer's (past) celestial sphere,
$(\alpha_{1},\alpha_{2}).$ The null geodesics of the past
lightcones from the observer's worldline thus carry these
labels; the points on each null geodesic are further labeled
by the parameter along the curve, which we take to be an
affine parameter $s$ suitably normalized so that
$l_{a}v^{a}=1$.  Thus the past lightcone of the observer has
the form 
\begin{equation}\label{pastlc} 
	x^{a}
   =
	X^{a}(x_{0}^{a}(\tau ),\alpha_{1},\alpha_2,s)  
\end{equation} 

\noindent where the functions
$X^{a}(x_{0}^{a}(\tau),\alpha_{1},\alpha_2,s)$ satisfy the
geodesic equation
\[ 
\dot{X}^{a}\nabla _{a}\dot{X}^{b}=0
\] 

\noindent with 
\[ 
	\dot{X}^{a}
  =
	\frac{\partial}{\partial s}
	X^{a}(x_{0}^{a}(\tau),\alpha_{1},\alpha_{2},s) 
\]

\noindent and the null condition

\begin{equation} \label{nullcondition}
	g_{ab}\dot{X}^{a}\dot{X}^{b}=0.   
\end{equation}

\noindent The local coordinates $x^{a}$ can be chosen so that
one of them, say $x^{0}$, is timelike and the remaining three
$x^{i}$ are spacelike, $i=1,2,3$.  In this case, the function
$\dot{X}^{0}$ does not vanish at any point.  Although we are
interested in the past lightcone, it is more straightforward
to work in terms of the future lightcone. The direction in
which the lightrays are traced is not important in the case
of interest, namely, static spacetimes.  Therefore,
$\dot{X}^0$ is everywhere positive.   

This means that $x^{0}$ increases monotonically with the
affine parameter $s$.  Because $\dot{X}^{0}$ does not vanish
anywhere, then, by (\ref {nullcondition}), at all points on a
geodesic one of $\dot{X}^{i}$ is non-zero (different $i$
possibly in different sections of the geodesic).  For
definiteness, we label this spatial coordinate by $i=1$.  From
the implicit function theorem, we have that

\begin{equation}
	x^{1}
   =
	X^{1}(x_{0}^{a}(\tau ),\alpha_{1},\alpha_{2},s) 
\end{equation}

\noindent can be inverted to obtain

\begin{equation}\label{parameter} 
	s
   =
	S(x_{0}(\tau ),\alpha_1,\alpha_2,x^{1}).  
\end{equation}

\noindent The inversion will only be possible in patches,
since it is possible that $\dot{X}^{1}$ vanishes at isolated
points.  This means that $s$ will be, in general, a
multiple-valued function of $x^{1}$.  Still, this inversion
allows us to reparametrize the geodesics in terms of just our
coordinates and observation angles:

\begin{eqnarray} 
	x^0 &=&X^0(\tau ,
		       \alpha_1,\alpha _2,
			S(\tau,\alpha_1,\alpha_2,x^1))
	  \equiv \hat{X}^0(\tau ,
			   \alpha_1,\alpha_2,
			   x^1), 	\label{timeofarr} \\
	x^1 &=&x^1 \label{radius} \\ 
	x^A &=&X^A(\tau,\alpha_1,\alpha_2,
			S(\tau,
			  \alpha_1,\alpha_2,x^1))
	\equiv \hat{X}^A(\tau,
			 \alpha_1,\alpha _2,x^1).\label{lenseq}
\end{eqnarray}

\noindent with $A=2,3$.  The idea is now to treat these
equations as if they determine a source at the spacetime
point\textbf{\ (}$x^{0},x^{1},x^{A})$ in terms of the
observable quantities\textbf{\ (}$\tau ,\alpha_1,\alpha
_{2})$ where we have assumed, {\it{for the moment}}, that the
coordinate value for $x^{1}$ can be determined from
observation.  We will treat the source as slow moving or
effectively at rest.  In this case, Eq.~(\ref{lenseq}) is
defined as the lens equation.

More specifically, we interpret this lens equation as
follows.  Consider a source at a spatial location $x^{i}$,
emitting light at time $x^{0}$.  We can think of the
coordinate $x^{1}$ as a type of radial coordinate.  The
remaining two coordinates $x^{A}$ are thus a type of angular
coordinates.  The emitted light arrives at the observer at a
time $\tau$, in a direction $(\alpha_1,\alpha_2)$. 
Eq.~(\ref{lenseq}) expresses the angular location of a source
at radial distance $x^{1}$ in terms of the observation angles
$(\alpha_1,\alpha_2)$.  On the other hand,
Eq.~(\ref{timeofarr}) is an exact ``time of arrival''
equation; it relates the time of emission, $x^{0}$, at the
radial location, $x^{1}$, with the observer's proper time
$\tau $, and arrival direction, $(\alpha_1,\alpha_2)$.

The lens equation, Eq.~(\ref{lenseq}), represents a map from
the image (or observation) angles $(\alpha_1,\alpha_2)$
to the source position angles, $x^{A}$.  The map breaks down
at locations where the determinant

\begin{equation} 
J(\tau ,\alpha_1,\alpha_2,x^{1})
\equiv \det \frac{\partial
(x^{2},x^{3})}{\partial (\alpha_1,\alpha_2)} \label{j} 
\end{equation}

\noindent vanishes.  $J(\tau ,\alpha_1,\alpha_2,x^{1})=0$
defines the caustics (a three surface in four-space) of the
family of past lightcones of the observer.  The direct
consequence of the break-down of the map is that multiple
images of the same source can be observed.  More specifically,
often one can see an image, in direction $(\alpha_1,\alpha
_{2})$, of an object that lies on a null geodesic before it
reaches a caustic (in affine distance), while simultaneously
seeing a different image, in a different direction
$(\alpha'_1,\alpha'_2)$, from the same object along a
different null geodesic, but, in this case, the object lies
beyond the caustic in affine distance.  The parity of the
images is given by the sign of $J$.  In background dependent
calculations, $J^{-1}$ is often interpreted as a magnification
factor with respect to the ``unlensed'' source, but, as we
have no background, this would not be appropriate here.  (Note
that $J(\tau ,\alpha_1,\alpha_2,x^1)$ could have been
calculated holding $s$ fixed instead of fixed $x^1;$ the
vanishing of $J$ is independent of that choice.  This follows
from the general theory of Lagrangian submanifolds and maps.)

The lens equation, Eq.~(\ref{lenseq}), is not yet entirely
usable since it involves the (up to now) unobservable quantity
$x^{1}$.  However, $x^{1}$ can be expressed in terms of
observable quantities through the use of the idea of
distance.  Though there are many definitions of distance in
use in general relativity and astrophysics, several of them
can be considered to be observable and we thus explore the
feasibility of inferring $x^{1}$ from the considerations of
distance.  We will investigate a definition of distance which
is observable, namely the so-called angular-diameter distance
- there being several closely related distance
definitions~\cite{EFN,Ehlers}.

Since we have, in principle, exact expressions for the past
lightcone of the observer in terms of parameters adapted to
the null geodesic congruence, we have a natural way of
expressing the angular-diameter distance to the source in
exact form.  The angular-diameter distance is
defined~\cite{Ehlers} in terms of the infinitesimal area
spanned by the observer's geodesic congruence at the location
of the source per infinitesimal solid angle at the observer's
location, namely:

\begin{equation} 
D_{A}=
\left| \frac{dA_{s}}{d\Omega _{0}}\right|^{\frac12}
\end{equation}

In order to calculate the area $dA_{s}$, we define two
connecting vectors in the lightcone of the observer.  By
taking variations of the points on the lightcone with respect
to the labels of the null geodesics in the congruence, we find
the geodesic deviation vectors, or Jacobi fields:

\begin{equation} 
M_{1}^{a}=\frac{\partial X^{a}}{\partial
\alpha_1},\hspace{1.5cm} M_{2}^{a}=\frac{\partial
X^{a}}{\partial \alpha_2}.  
\end{equation}

\noindent It is irrelevant to the area calculation whether
$s$\ or $x^{1}$\ are held constant in calculating the
connecting vectors -- the difference, lying along the null
tangent vectors to the geodesics, does not affect the area.

The area $dA_{s}$ is the area spanned by these two vectors at
the location of the source, namely, the norm of the
wedge-product of the two vectors:

\begin{eqnarray}   
dA_{s}&=&\left|
g_{ac}g_{bd}M_{1}^{[a}M_{2}^{b]}M_{1}^{[c}M_{2}^{d]}\right|^{\frac12}\,d\alpha
_{1}\,d\alpha_2 \nonumber \\ 
&=&\left| 2\left((M_{1}\!\cdot
\!M_{1})\,(M_{2}\!\cdot \!M_{2})-(M_{1}\!\cdot
\!M_{2})^{2}\right)\right| ^{\frac{1}{2 }}\,d\alpha_1\,d\alpha_2.
\label{area}  
\end{eqnarray}

\noindent If the solid angle at the observer subtended by the
area $dA_s$ is given by $d\Omega _{0}=K
(\alpha_1,\alpha_2)\,d\alpha_1\,d\alpha_2$, where $K$ depends
on the choice of the coordinates, the angular-diameter
distance is given by

\begin{equation} D_A^2 = D_A^2 
(\alpha_1, \alpha_2, \tau, x_1) = 2\,K^{-2}\,\left|
(M_{1}\!\cdot \!M_{1})\,(M_{2}\!\cdot \!M_{2})-
(M_{1}\!\cdot \!M_{2})^{2})\right|.
\label{DA} \end{equation}

\noindent For sufficiently small values of $x^{1}$,
Eq.~(\ref{DA}) is invertible, i.e.,
$x^{1}=x^{1}(\alpha_1,\alpha_2, \tau, D_{A})$.  However,
$D_{A}$ goes to zero at the caustic, so that beyond the
caustic $x^{1}$ is a multivalued function of $D_{A}$ and must
be given in patches.

The angular-diameter distance is observable, because it is
related to the intrinsic luminosity $L$ of the source and its
apparent brightness $S$ (total flux at the observer) via~\cite
{Ehlers}

\begin{equation} 
S=\frac{L}{4\pi (1+z)^{4}D_{A}^{2}}.  
\label{candle} 
\end{equation}

\noindent In principle, Eq.~(\ref{candle}), with
Eq.~(\ref{DA}), gives $x^{1}$ implicitly as a function of
observables:  the angular location of the image
$(\alpha_{1},\alpha _{2})$, its redshift $z
(=\omega_s/\omega_0-1)$ its apparent brightness and the
intrinsic luminosity of the source.

On the other hand, there may be situations where the intrinsic
luminosity of the source is not available.  In such cases, if
there are multiple images observed, then we can make use of
their relative brightness in order to estimate $x^{1}$.  For
two images of the same source, lying at angles
$(\alpha_1^{(1)},\alpha_2^{(1)})$ and $(\alpha
_{1}^{(2)},\alpha_2^{(2)})$, the ratio of the fluxes
$S_{1}/S_{2}$ does not depend on the intrinsic luminosity of
the source $L$ and can be interpreted as the relative
magnification $\mu _{12}$ of one image with respect to the
other one, or

\begin{equation}  
\mu
_{12}=\frac{D_{A}^{2}(\alpha_1^{(2)},\alpha_2^{(2)},x^{1})}{
D_{A}^{2}(\alpha_1^{(1)},\alpha_2^{(1)},x^{1})}. 
\label{murel} 
\end{equation}

\noindent Notice that $(\alpha_1^{(1)},\alpha_2^{(1)})$ and
$(\alpha _{1}^{(2)},\alpha_2^{(2)})$ are two image directions
of Eq.~(\ref{lenseq}) for a given value of the source
coordinates $(x^{2},x^{3})$.  The inversion of
Eq.~(\ref{murel}) is not likely to be feasible in closed
form.  Still, in principle, Eq.~(\ref {murel}) gives $x^{1}$
implicitly in terms of the angular location of two images and 
their relative brightness $\mu _{12}\equiv S_{1}/S_{2}$.

(In the case of lensing at cosmological distances, it is
customary to infer distances from redshifts.  Even though we
are not concerned with cosmological models in this paper, we
consider redshifts as another alternative to infer $x^{1}$
from an observable quantity. For a source on a worldline with
tangent vector $v_{s}^{a}$, emitted light of frequency $\omega
_{s}$ and observed frequency $\omega _{0},$ the ratio $\omega
_{s}/\omega _{0}$ is given by~\cite{Schrodinger}

\begin{equation} \frac{\omega _{s}}{\omega
_{0}}=\frac{g_{ab}(x^{a}(\alpha_1,\alpha
_{2},x^{1}))v_{s}^{a}\dot{X}^{b}(\alpha_1,\alpha_2,x^{1})}{
g_{ab}(x_{0}^{a})v_{0}^{a}\dot{X}^{b}(x_{0}^{a})}
\label{omega} 
\end{equation}

\noindent where Eq.~(\ref{lenseq}) has been used for the
source's space-time location. Equation~(\ref{omega}) gives
$x^{1}$ implicitly in terms of the frequency of emission, the
received frequency and the observed image angle.  As our
assumed source is at rest, its velocity is $v_{s}^{a} =
|g_{00}(x^a(\alpha_1,\alpha_2,x^1))|^{-1/2} (1,0,0,0)$.)

\section{The Schwarzschild Case}

\subsection{The Lens Equation}

We consider now the case of gravitational lensing by a
Schwarzschild black hole of mass $M$.  The line element is

\begin{equation}
ds^{2}=f(r)dt^{2}-\frac{1}{f(r)}dr^{2}-r^{2}(d\theta ^{2}+\sin
^{2}\!\theta d\phi ^{2}) 
\label{line} 
\end{equation}
with
\begin{equation} 
f(r)\equiv 1-\frac{2M}{r} 
\end{equation}

\noindent In order to take advantage of existing
calculations~\cite{kling}, we temporarily use coordinates
$(u,l)$ given by

\begin{eqnarray} u &=&\frac{1}{\sqrt{2}}\left( t-\int
\frac{dr}{f}\right) =\frac{1}{\sqrt{2}} \left(t-r+2M\log
(2M-r) \right), \\ l &=&\frac{1}{\sqrt{2}r}. 
\end{eqnarray}

\noindent In these coordinates, the line element takes the
form

\begin{equation} ds^{2}= 2f\,du^{2} -\frac{2}{l^{2}}du\,dl
-\frac{1}{2l^{2}}(d\theta ^{2}+\sin ^{2}\!\theta \,
d\phi^{2}).   
\end{equation}

\noindent The equations for \textit{null\/} geodesics
$\ddot{x}^a+\Gamma _{bc}^a \dot{x}^b\dot{x}^c=0$ in terms of
an affine parameter $s$ are equivalent to

\begin{eqnarray} \dot{u} &=&\frac{C}{2f} \left( 1\pm
\sqrt{1-\left( \frac{B}{C}\right) ^{2}l^{2}f}\right)
\label{udot} \\ 
\dot{l} &=&\pm Cl^{2}\sqrt{1-\left(
\frac{B}{C}\right) ^{2}l^{2}f} \label{ldot} \\ 
\dot{\phi}
&=&\frac{Al^{2}}{\sin ^{2}\theta } \label{phidot} \\ 
\left(
\frac{\dot{\theta}}{l^{2}}\right) ^{2}
&=&B^{2}-\frac{A^{2}}{\sin ^{2}\theta } \label{thetadot}
\end{eqnarray}

\noindent with the null condition $\dot{x}^{a}\dot{x}_{a}=0$
equivalent to

\begin{equation} 4l^{2}f\dot{u}^{2}-4Cl^{2}\dot{u}
+\dot{\theta}^{2}+\frac{A^{2}l^{4}}{\sin ^{2}\theta }=0. 
\end{equation}

\noindent The symbol ($\dot{\;})\ $stands for $d/ds$ and
$A,B,C$ are three first integrals of the null geodesics,
depending on the initial point and the initial direction.  The
constant $C$ represents the freedom in the scaling of the
affine parameter $s$.  Treating the observers location
$(u_{0},l_{0},\theta _{0},\phi _{0})$ as the initial point,
the constant $B$ is related to the angle $\psi $ that the null
geodesic makes with the optical axis (defined by the radial
line from observer to the lens center).  More precisely:

\begin{equation} 
\frac{B}{C}=\frac{\sin \psi }
	{l_{0}\sqrt{f(l_{0})}}, \label{const1}
\end{equation}

\noindent where $l_{0}$ is the inverse radial location of the
observer.  Lastly, the constant $A$ can be related to the
azimuthal angle $\gamma $ that the direction of the lightray
makes around the optical axis at the observer's location via

\begin{equation} \frac{A}{C} = \sin \theta _{0}\sin \gamma
\frac{\sin \psi }{l_{0}\sqrt{f(l_{0}) }}, \label{const2}
\end{equation}

\noindent where $\theta _{0}$ is the angular location of the
observer.

We can switch from the null coordinate $u$ to the time
coordinate $t$ using $\dot{t}=\sqrt{2}\left(
\dot{u}-\frac{1}{2l^{2}f}\dot{l}\right)$.  Doing so and
setting $C =1$ allows Eqs.~(\ref{udot}-\ref{thetadot}) to be
rewritten as

\begin{eqnarray} \dot{t} &=&\frac{1}{\sqrt{2}f}
\label{timedot2} \\ \dot{l} &=&\pm l^{2}\sqrt{1-\left(
\frac{\sin ^{2}\psi }{l_{0}^{2}f(l_{0})} \right) l^{2}f}
\label{ldot2} \\ \dot{\phi} &=&\frac{\sin \theta _{0}\sin
\gamma \sin \psi l^{2}}{l_{0}\sqrt{ f(l_{0})}\sin ^{2}\theta }
\label{phidot2} \\ \left( \frac{\dot{\theta}}{l^{2}}\right)
^{2} &=&\frac{\sin ^{2}\psi }{ l_{0}^{2}f(l_{0})}-\frac{(\sin
\theta _{0}\sin \gamma \sin \psi )^{2}}{ l_{0}^{2}f(l_{0})\sin
^{2}\theta } \label{thetadot2} 
\end{eqnarray}

For those null geodesics of interest to us, i.e., those whose
initial direction has a component pointing towards the
Schwarzschild origin, the inverse radial distance $l$
initially increases ($\dot{l}>0$) until the point of closest
approach to the lens is reached, (with affine parameter value
$s_{p})$.  The coordinate, $l$, then decreases ($\dot{l}<0$)
after $s_{p}$ until reaching the source at some $s_{fin}$. 
However, for $s<s_{p}$ and for $s>s_{p}$ we have $\dot{l}\neq
0$, and thus the inverse radial distance $l$ can be used as a
parameter (in two patches, the incoming and outgoing) along
the null geodesics for the purposes of constructing our lens
equation in the manner of the previous Section.  It plays the
role of $x^{1}.$

\begin{figure}
\centerline{\psfig{figure=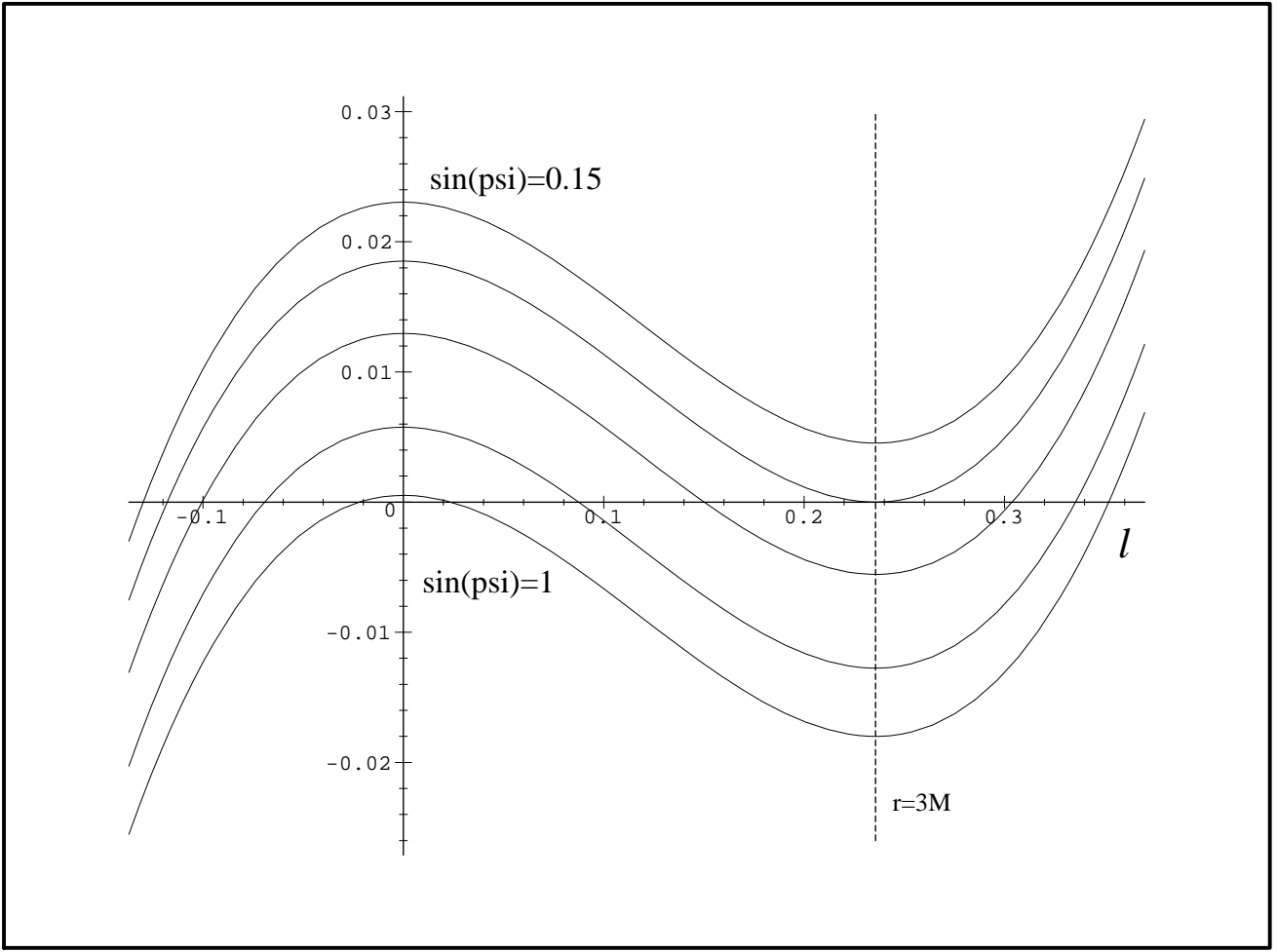,width=7.5in,angle=-90}}

\caption{A plot of the function $F:=1-\sin^2\!\psi
l^2f/(l_0^2f(l_0))$ as a function of $l$ for a sequence of
values of $\sin\psi$ including $\sin\psi=1$ .  Notice that for
large enough $\psi$ there are always two positive roots, while
for small enough $\psi$ there are no positive roots.  There is
a critical value of $\psi$ for which there is only one
positive root.  The allowed values of $l$ are such that $F$ is
positive, and such that $l=0$ is included.  Thus, allowed
values of $l$ range from 0 to the smallest positive root. The
smallest positive root is denoted $l_p$ and represents the
point of closest approach to the lens.  Notice that $l_p$
increases with decreasing $\psi$, reaching the critical value
$(3\protect\sqrt{2}M)^{-1}$ at the critical value of $\psi$. 
} \label{l0} \end{figure}

The value of $l$ at the point of closest approach (i.e.  at
$\dot{l} =0$) is denoted $l_{p}.$ If we assume, naturally,
that the observer is located outside the last stable orbit (at
$r>3M$), then, for lightrays that do not cross the $r=3M$
orbit, the closest approach $l_{p}$ is the smallest of the
positive roots of \begin{equation} {(\sin \psi
)^{2}}~l_{p}^{2}~(1-2\sqrt{2}Ml_{p})-l_{0}^{2}~(1-2\sqrt{2}
M~l_{0})=0.  \label{lp} \end{equation} A simple analysis shows
that for $l_{0}<3\sqrt{2}M$, Eq.~(\ref{lp}) has no positive
roots unless

\begin{equation} \sin \psi > 3\sqrt{2}Ml_{0}
\sqrt{3(1-2\sqrt{2}Ml_{0})} \label{psilimit} \end{equation}

\noindent in which case there are always two positive roots,
and the closest approach $ l_{p}$ is the smallest of them.  It
is simple to prove that $l_{p}<(3\sqrt{2} M)^{-1}$ for all
$\psi $ subject to Eq.~(\ref{psilimit}), and that $l_{p}\to (3
\sqrt{2}M)^{-1}$ for $\sin \psi \to
3\sqrt{2}Ml_{0}\sqrt{3(1-2\sqrt{2}Ml_{0}) }$.  See
Fig.~\ref{l0}.

The $l_{p},$ which is a turning point of the coordinate $l$
along the null geodesics, plays a major role in the
following.  First we notice that the term,

\[ 1-\left( \frac{\sin ^{2}\psi} {l_{0}^{2}f(l_{0})}\right)
l^{2}f \]

\noindent from the $\dot u$ equation, can be rewritten with
the role of $\psi $ now played by $l_{p},$ in the form

\begin{equation} 1-\left( 
\frac{\sin ^{2}\psi }{l_{0}^{2}f(l_{0})}\right) l^{2}f
=\frac{1}{
l_{p}^{2}(1-2\sqrt{2}Ml_{p})}
\bigg(l_{p}^{2}(1-2\sqrt{2}Ml_{p})-l^{2}(1-2 \sqrt{2}Ml)\bigg)
\label{udot3} \end{equation}

\noindent using, from Eq.~(\ref{lp}),

\begin{equation} \frac{\sin ^{2}\!\psi
}{l_{0}^{2}(1-2\sqrt{2}Ml_{0})}=\frac{1}{l_{p}^{2}(1-2
\sqrt{2}Ml_{p})}.  \end{equation}

\noindent We thus see that the dependence on $\psi $ is now
hidden away in the inverse radial distance of closest approach
$l_{p}$.  This observation will simplify some of the
calculations that follow.

The past lightcone of an observer in coordinates
$x^{a}=(t,l,\theta ,\phi )$ in terms of the affine parameter
$s$ and initial directions $(\psi ,\gamma )$ could in
principle be obtained from Eq.~(\ref{udot} - \ref{thetadot})
or Eq.~(\ref{timedot2} - \ref{thetadot2}). This requires the
integration of four non-linear ordinary differential equations
which can not be done by quadratures.  In the spirit of the
Section II, however, we do not need the lightcone in terms of
the affine parameter $s$, but in terms of a radial
coordinate.  Our radial coordinate is the inverse radial
distance $l$, which is better suited for treating large
distances than the standard $r$.  In particular, the infinite
range $3M<r<\infty $ translates into the finite interval
$0<l<(3\sqrt{2}M)^{-1}$.

First we show how our radial coordinate $l$ is related to the
affine length.  Next, we integrate the lightcone in terms of
$l$.  For our purposes, it suffices to assume that by the time
the lightray reaches the observer it has already passed by the
point of closest approach $l_{p}$.

Eq.~(\ref{ldot2}) can be integrated to obtain the affine
parameter $s$ in terms of the inverse radial distance:

\begin{eqnarray} 
	s&=&2\int_{l_{0}}^{l_{p}}
	\sqrt{\frac{l_p^2(1-2\sqrt{2}Ml_p)}
		   {l_p^2(1-2\sqrt{2}Ml_{p})
		    -l'^2(1-2\sqrt{2}Ml')}
	      } 
	\frac{dl'}{l'^2} \nonumber \\ 
	& & ~+\int_l^{l_{0}}
	\sqrt{\frac{l_p^2(1-2\sqrt{2}Ml_p)}
	    	   {l_p^2(1-2\sqrt{2}Ml_p)
		    -l'^2(1-2\sqrt{2}Ml')}}
	\frac{dl'}{l'^2} .\label{s} 
\end{eqnarray}

Equation~(\ref{s}) corresponds to Eq.~(\ref{parameter}) of the
previous section.  This represents one of the two available
patches for $s$ as a function of $l.$

\begin{figure}
\centerline{\psfig{figure=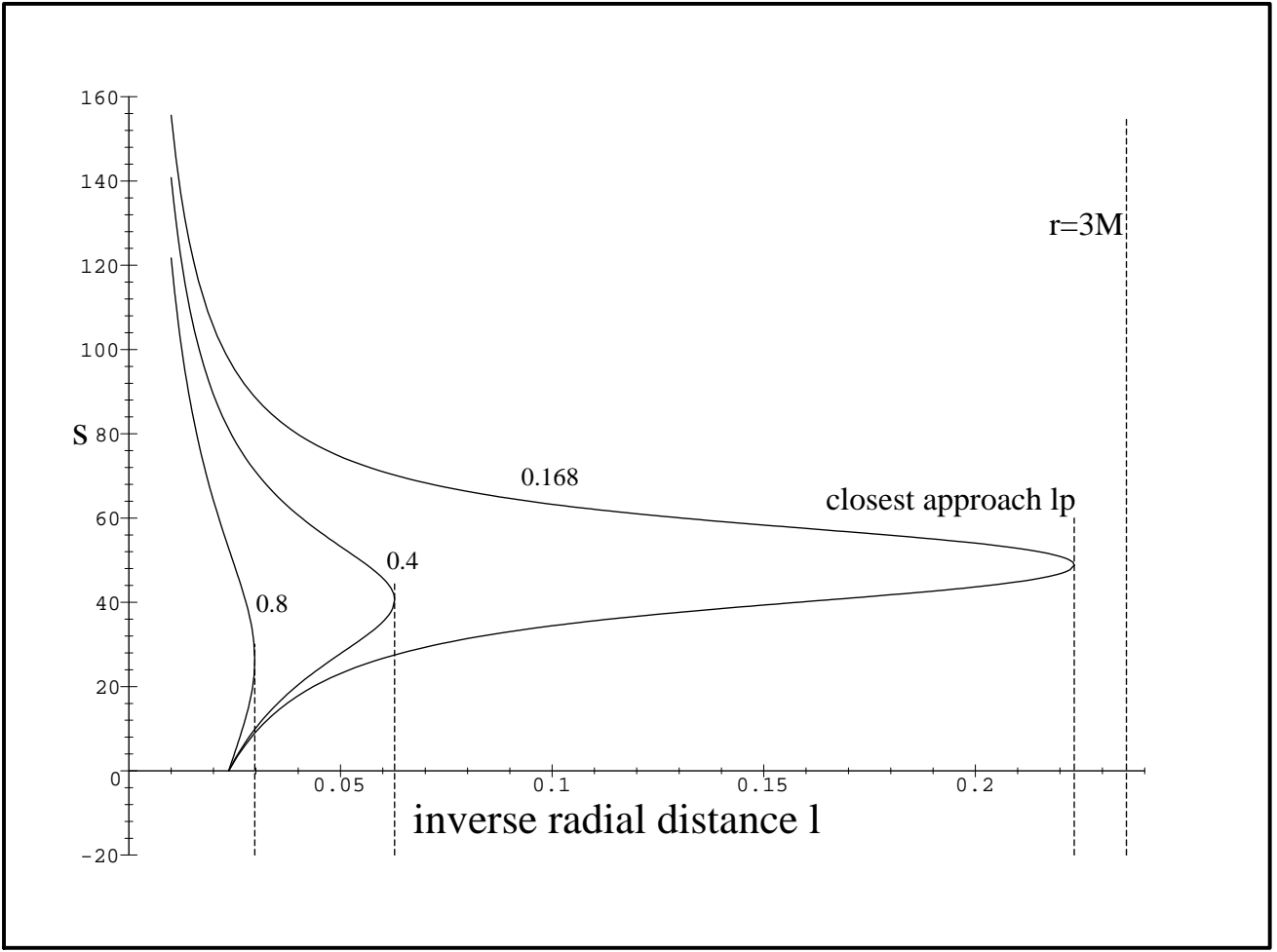,width=7.5in,angle=-90}}

\caption{The affine length $s$ as a function of the inverse
radial distance l for three null geodesics that reach the
observer from positive directions $ \psi$.  We have set the
observer at $l_0=0.1/3\protect\sqrt{2}M$, the mass is $M=1$,
and the constant $C$ has the value 1 for the three null
geodesics.  The three null geodesics are labeled according to
the value of $\sin\psi$.  Smaller positive angles $\psi$ reach
closer to the $3M$ radius at their respective point of closest
approach $l_p$.  For this observer the smallest image angle is
at $\sin\psi=0.16733201$, at which the closest approach
reaches the $3M$ radius.  We see that the affine length is
double valued as a function of the inverse radial distance.}
\label{sfigure} \end{figure}

The affine parameter as a function of $l$ is represented in
Fig.~\ref{sfigure} when the observer is at a distance of
$30M$.  (We chose a relatively small distance in order to
better appreciate the strong field effects.)  The affine
length goes to infinity as $l$ approaches zero, in agreement
with the fact that the lightray runs out to infinity.  The
affine length, $s$, is chosen to vanish at the observer's
location.  We see, in the diagram, that the affine length
bulges towards the $3M$ radius, resulting in a double valued
function of $l$.  The bulge is more pronounced for lightrays
that reach the observer at smaller observation angles $\psi
$.  The rays that come closer to the $3M$ radius spend more
affine time in reaching the observer, in agreement with the
gravitational time delay.

In the following, we obtain the past lightcone of the observer
$(t_0, l_0, \theta_0, \phi_0)$ as a function of the inverse
radial distance $l$, instead of the affine parameter $s$, and
two angles $(\psi, \gamma)$ specifying the direction of each
null geodesic at the observer's location.

Integrating Eq.~(\ref{timedot2}) with Eq.~(\ref{ldot2}), we
obtain

\begin{eqnarray} t &=&t_{0}+2\int_{l_{0}}^{l_{p}}
\sqrt{\frac{l_{p}^{2}(1-2\sqrt{2}Ml_{p})}{
l_{p}^{2}(1-2\sqrt{2}Ml_{p})-l^{2}
(1-2\sqrt{2}Ml)}}\frac{dl^{\prime }}{\sqrt{ 2}l^{\prime
}{}^{2}(1-2\sqrt{2}Ml^{\prime })} \nonumber \\
&&+\int_{l}^{l_{0}}
\sqrt{\frac{l_{p}^{2}(1-2\sqrt{2}Ml_{p})}{l_{p}^{2}(1-2
\sqrt{2}Ml_{p})-l^{2}(1-2\sqrt{2}Ml)}}\frac{dl^{\prime
}}{\sqrt{2}l^{\prime }{}^{2}(1-2\sqrt{2}Ml^{\prime })}
\label{t} 
\end{eqnarray}

\noindent This is the equivalent of Eq.~(\ref{timeofarr}) of
the previous section.  As a function of $l$, the time of
arrival is double-valued (not so as a function of the affine
parameter $s$); Eq.~(\ref{t}) represents one of the two
patches.

The integration of the angular coordinates of the lightcone is
carried out in~\cite{kling}. Representing the angular
coordinates $(\theta ,\phi )$ in terms of the complex
stereographic variables $\zeta \equiv \cot (\theta
/2)e^{i\phi}$ the integration yields

\begin{equation} \zeta =e^{i\phi _{0}}\frac{\cot
\frac{\theta_0}{2}+e^{i\gamma }\cot \frac{ \Theta
(l,l_0,l_p)}{2}}{1-e^{i\gamma }\cot \frac{\Theta
(l,l_0,l_p)}{2}\cot \frac{\theta_0}{2}} \label{zeta}
\end{equation}

\noindent where $\Theta (l,l_{0},l_{p})$ is

\begin{eqnarray} \Theta (l,l_{0},l_{p}) &=& \pm\left(\pi -
2\int_{l_{0}}^{l_{p}}\frac{dl}{\sqrt{
l_{p}^{2}(1-2\sqrt{2}Ml_{p})-l^{2}(1-2\sqrt{2}Ml)}} \right. 
\nonumber \\ &~&\quad \left.  -\int_{l}^{l_{0}}\frac{
dl^{\prime }}{\sqrt{l_{p}^{2}(1-2\sqrt{2}Ml_{p})-l^{\prime
}{}^{2}(1-2\sqrt{2 }Ml^{\prime })}} \right).  \label{Theta}
\end{eqnarray}

\noindent The function $\Theta (l,l_{0},l_{p})$ depends on the
observation angle $\psi $ through $l_{p}$.  The overall
positive sign is taken when the value of the observation
angle, $\psi$, is positive, and the negative sign is taken for
negative $\psi$.  This makes $\Theta (l,l_{0},l_{p})$ an odd
function of $\psi$.

Geometrically, $\Theta (l,l_0, l_p)$ ``represents'' the
angular position of the source relative to the optical axis,
defined by the line between the lens and observer.  The
observer is considered to lie on the optical axis at $\Theta =
\pi$.  The relative angular position of a source is given by
$\Theta$ values between $-\pi$ and $\pi$.  Hence,
Eq.~(\ref{Theta}) must be considered mod $2\pi$, where values
outside the range, $-\pi \le \Theta \le \pi$, represent
multiple circlings of the lens.  When $\Theta = 0, 2\pi,
4\pi,$ \ldots, the source is colinear with the lens and
observer and would be observed as an Einstein Ring.  For
positive $\psi$, a value of $\Theta$ mod $2\pi$ between $\pi$
and $0$ represents a source located to the right of the lens,
while $-\pi < \Theta~{\rm{mod}}~2\pi< 0$ represents a source
located to the left of the optical axis.

Figure~\ref{lenseqfigure} shows a plot of $\Theta $ at fixed
values of $l$ and $l_{0}$, as a function of the image angle
$\psi $.  We can see that $\Theta $ blows up at
$l_{p}=(3\sqrt{2}M)^{-1}$, which agrees with the fact that, as
the lightrays approach the $3M$ radius, they take a larger
number of turns around the lens.  Notice that $\Theta $ is a
regular function of $l$ for all $l<l_{p}$ because the
integrand diverges slowly, as $ (l_{p}-l)^{-1/2} $.  In fact,
for numerical integration it turns out to be much more
efficient to make a change of variables $l=l_{p}-q$ and write
$\Theta $ as

\begin{eqnarray} \Theta &=&\pm\left(\pi
-2\int_{0}^{l_{p}-l_{0}}\frac{dq}{\sqrt{2l_{p}(1-2\sqrt{2}
Ml_{p})q+(6\sqrt{2}Ml_{p}-1)q^{2}-2\sqrt{2}Mq^{3}}} \right. 
\nonumber \\ &~& \quad\left. -\int_{l}^{l_{0}}\frac{
dl^{\prime }}{\sqrt{l_{p}^{2}(1-2\sqrt{2}Ml_{p})-l^{\prime
}{}^{2}(1-2\sqrt{2 }Ml^{\prime })}}\right).  \label{theta4}
\end{eqnarray}

In terms of the standard spherical coordinates $(\theta ,\phi
)$, Eq.~(\ref{zeta}) translates into:

\begin{eqnarray} \cos \theta &=&-\cos \theta _{0}\cos \Theta
+\sin \theta _{0}\sin \Theta \cos \gamma \label{theta}
\nonumber \\ \tan \phi &=&\frac{\sin \phi _{0}\sin \theta
_{0}-\tan \Theta \left( \cos \phi _{0}\sin \gamma -\sin \phi
_{0}\cos \gamma \cos \theta _{0}\right) }{ \cos \phi _{0}\sin
\theta _{0}+\tan \Theta \left( \sin \phi _{0}\sin \gamma +\cos
\phi _{0}\cos \gamma \cos \theta _{0}\right) }\label{thetaphi}
\end{eqnarray}

\noindent Equations (\ref{thetaphi}), with Eq.~(\ref{theta4})
are the exact lens equations for the case of a Schwarzschild
spacetime and correspond to Eqs.~(\ref {lenseq}) of the
previous section.  Notice that the observer is located at
generic values of $(\theta _{0},\phi _{0})$, which means that
we do not choose,\ as is often done\textbf{,} the $z-$axis as
the optical axis, the optical axis being the radial line that
contains both the center of symmetry and the observer.  This
is because the spherical coordinates break down along the
$z-$axis.  If we chose the observer to lie along the $z-$
axis, then Eq.~(\ref{thetaphi}) reduce to $\cos \theta =\cos
\Theta $ and $\tan \phi =\tan \gamma $, and we could interpret
$\Theta $ and $\gamma $ as the lens angular coordinates. 
However, this would result in erroneous predictions in the
following subsections, unless we use additional care.  In
order to keep the remainder of this paper in the most
transparent form, we prefer to keep the observer off the
$z-$axis.

\subsection{Lensing Observables}

In this subsection, we describe the calculation of three key
lensing observables from the lens equations:  the
angular-diameter distance, the relative magnifications, and
the time-delay between the arrival times of two images. 

We start by exploring the angular-diameter distance, using
Eqs.~(\ref{thetaphi}) and Eq.~(\ref{t}) to obtain an exact
expression of the angular-diameter distance in terms of the
inverse parameter $l$.  In the next subsection, we will use the
expressions we obtain here to explore the possibility of
inferring the inverse radial distance, $l$, to the source.

First, we define the connecting vectors

\begin{eqnarray} M_{1}^{a} &\equiv &\left( \frac{\partial
t}{\partial \gamma },\frac{\partial l}{\partial \gamma
},\frac{\partial \theta }{\partial \gamma },\frac{ \partial
\phi }{\partial \gamma }\right) =\left( 0,0,\frac{\partial
\theta }{ \partial \gamma },\frac{\partial \phi }{\partial
\gamma }\right) \\ M_{2}^{a} &\equiv &\left( \frac{\partial
t}{\partial \psi },\frac{\partial l }{\partial \psi
},\frac{\partial \theta }{\partial \psi },\frac{\partial \phi
}{\partial \psi }\right) =\left( \frac{\partial t}{\partial
\psi },0, \frac{\partial \theta }{\partial \psi
},\frac{\partial \phi }{\partial \psi } \right) 
\end{eqnarray}

\noindent where $t,\theta ,\phi $ are functions of $(l,\psi
,\gamma )$ given by Eq.~(\ref{t}) and Eq.~(\ref{thetaphi}). 
The partial derivatives are taken at fixed value of $l$.  From
the expressions above, we have

\begin{eqnarray} \frac{\partial \theta }{\partial \gamma }
&=&\frac{\sin \Theta \sin \theta _{0}\sin \gamma
}{\sqrt{1-(\cos \Theta \cos \theta _{0}-\sin \Theta \sin
\theta _{0}\cos \gamma )^{2}}} \label{connectinga} \\
\frac{\partial \theta }{\partial \psi } &=&-\frac{\sin \Theta
\cos \theta _{0}+\cos \Theta \sin \theta _{0}\cos \gamma
}{\sqrt{1-(\cos \Theta \cos \theta _{0}-\sin \Theta \sin
\theta _{0}\cos \gamma )^{2}}}\frac{\partial \Theta }{\partial
\psi } \label{connectingb} \\ \frac{\partial \phi }{\partial
\gamma } &=&-\frac{\sin \Theta (\sin \Theta \cos \theta
_{0}+\cos \Theta \sin \theta _{0}\cos \gamma )}{1-(\cos \Theta
\cos \theta _{0}-\sin \Theta \sin \theta _{0}\cos \gamma
)^{2}} \label{connectingc} \\ \frac{\partial \phi }{\partial
\psi } &=&-\frac{\sin \theta _{0}\sin \gamma }{1-(\cos \Theta
\cos \theta _{0}-\sin \Theta \sin \theta _{0}\cos \gamma
)^{2}}\frac{\partial \Theta }{\partial \psi }
\label{connectingd} \end{eqnarray}

Notice that, by Eq.~(\ref{connectinga}) and
Eq.~(\ref{connectingc}), the vector $ M_{1}^{a}$ is
proportional to $\sin \Theta $ for all generic values of $
\theta _{0}$ except for $\theta _{0}=0,\pi $.  This means
that, generically, the vector $M_{1}^{a}$ vanishes at $\Theta
=0$, which, by Eq.~(\ref{thetaphi}), represents source points
$(\theta ,\phi )$ along the optical axis.  If we had chosen
the optical axis as the $z-$axis this essential fact would not
be as transparent.

With the metric, Eq.~(\ref{line}), we have

\begin{eqnarray} M_{1}\cdot M_{1} &=&-\frac{1}{2l^{2}}\left(
\left( \frac{\partial \theta }{ \partial \gamma }\right)
^{2}+\sin ^{2}\!\theta \left( \frac{\partial \phi }{ \partial
\gamma }\right) ^{2}\right)
						\label{M1sq}\\ 
M_{2}\cdot M_{2} &=&f\left( \frac{\partial
t}{\partial \psi }\right) ^{2}- \frac{1}{2l^{2}}\left( \left(
\frac{\partial \theta }{\partial \psi }\right) ^{2}+\sin
^{2}\!\theta \left( \frac{\partial \phi }{\partial \psi
}\right) ^{2}\right) 
						\label{M2sq}\\
M_{1}\cdot M_{2} &=&-\frac{1}{2l^{2}}\left( \frac{\partial
\theta }{\partial \gamma }\frac{\partial \theta }{\partial
\psi }+\sin ^{2}\!\theta \frac{ \partial \phi }{\partial
\gamma }\frac{\partial \phi }{\partial \psi } \right)
\label{M1M2}. \end{eqnarray}

\noindent 
Using Eqs.~(\ref{M1sq} - \ref{M1M2}) the area $dA_{s}$ from
Eq.~(\ref{area}) can be written as

\begin{equation} 
   	dA_{s}
   =
	\left( \frac{\sin^2\!\theta}{4l^{4}}
	  \left(\frac{\partial\theta}{\partial\gamma}
		\frac{\partial\phi}{\partial\psi}
	       -\frac{\partial\phi}{\partial\gamma}
		\frac{\partial\theta}{\partial\psi}
	  \right)^2
	+f\left(\frac{\partial t}{\partial\psi}\right)^2
	  M_{1}\cdot M_{1}
	\right)^\frac12 d\psi d\gamma .  \label{area1} 
\end{equation}

\noindent The determinant of the lens map $J$ (Eq.~(\ref{j})),

\[ J=\frac{\partial \theta }{\partial \gamma }\frac{\partial
\phi }{\partial \psi }-\frac{\partial \phi }{\partial \gamma
}\frac{\partial \theta }{ \partial \psi } \]

\noindent which appears in Eq.~(\ref{area1}) can be simplified
using Eqs.~(\ref{connectinga} - \ref{connectingd}):

\begin{equation} J = \frac{\sin \Theta }{\sqrt{1-(\cos \Theta
\cos \theta _{0}-\sin \Theta \sin \theta _{0}\cos \gamma
)^{2}}}\frac{\partial \Theta }{ \partial \psi }.  \label{J}
\end{equation}
The scalar product $M_1\cdot M_1$ can also be evaluated using
Eqs.~(\ref{connectinga} - \ref{connectingd}):
\begin{equation}
M_1\cdot M_1  =  -\frac{\sin^2\Theta}{2l^2}
\end{equation}

\noindent Thus Eq.~(\ref{area1}) becomes

\begin{equation} dA_{s}=\frac{\sin \Theta }{2l^{2}}\left(
\left( \frac{\partial \Theta }{ \partial \psi }\right)
^{2}-2l^{2}f\left( \frac{\partial t}{\partial \psi } \right)
^{2}\right) ^{\frac{1}{2}}d\psi d\gamma \label{area2}
\end{equation}

\noindent The solid angle at the observer's location is
$d\Omega _{0}=\sin \psi\, d\psi \, d\gamma $.  The
angular-diameter distance is thus

\begin{equation} D_{A}^{2}=\frac{\sin \Theta }{2l^{2}\sin \psi
}\left( \left( \frac{\partial \Theta }{\partial \psi }\right)
^{2}-2l^{2}f\left( \frac{\partial t}{ \partial \psi }\right)
^{2}\right) ^{\frac{1}{2}} \label{D1} \end{equation}

This expression for the angular-diameter distance can be
simplified by showing that $\partial t/\partial \psi $ can be
expressed as a linear function of $ \partial \Theta /\partial
\psi $.  We present a short, intuitive derivation of this fact
here; a more formal derivation is given in the appendix.

First, we notice that both connecting vectors $M_{1}^{a}$ and
$M_{2}^{a}$ lie on the lightcone and therefore must be
orthogonal to the null vector that is tangent to the
lightrays, with components given by $\ell ^{a}\equiv
(\dot{t},\dot{l},\dot{\theta},\dot{ \phi})$.  The scalar
product of $M_{2}^{a}$ with $\ell ^{a}$ is

\begin{equation} g_{ab}M_{2}^{a}\ell
^{b}=f\dot{t}\frac{\partial t}{\partial \psi }-\frac{1}{
2l^{2}}\left( \dot{\theta}\frac{\partial \theta }{\partial
\psi }+\sin ^{2}\theta \dot{\phi}\frac{\partial \phi
}{\partial \psi }\right) \label{preprop} \end{equation}

\noindent where $\dot{\theta}$ and $\dot{\phi}$ are explicitly
given by

\begin{eqnarray} \dot{\theta} &=&\frac{\sin \Theta \cos \theta
_{0}+\cos \Theta \sin \theta _{0}\cos \gamma }{\sqrt{1-(\cos
\Theta \cos \theta _{0}-\sin \Theta \sin \theta _{0}\cos
\gamma )^{2}}}\frac{\sin \psi }{\sqrt{l_{0}^{2}(1-2\sqrt{2}
Ml_{0})}} \nonumber \\ \dot{\phi} &=&\frac{\sin \theta
_{0}\sin \gamma }{1-(\cos \Theta \cos \theta _{0}-\sin \Theta
\sin \theta _{0}\cos \gamma )^{2}}\frac{\sin \psi }{\sqrt{
l_{0}^{2}(1-2\sqrt{2}Ml_{0})}} \label{angledots}\end{eqnarray}

By inserting Eq.~(\ref{connectingb}), Eq.~(\ref{connectingd}),
and Eqs.~(\ref{angledots}) into Eq.~(\ref{preprop}), we have

\begin{equation} g_{ab}M_{2}^{a}\ell
^{b}=\frac{1}{\sqrt{2}}\frac{\partial t}{\partial \psi }+
\frac{\sin \psi
}{2\sqrt{l_{0}^{2}(1-2\sqrt{2}Ml_{0})}}\frac{\partial \Theta
}{\partial \psi }.  \end{equation}

\noindent Since $g_{ab}M_{2}^{a}\ell ^{b}=0$, we obtain the
claimed result

\begin{equation} \frac{\partial t}{\partial \psi }=-\frac{\sin
\psi }{\sqrt{2l_{0}^{2}(1-2 \sqrt{2}Ml_{0})}}\frac{\partial
\Theta }{\partial \psi }.  \label{relThetat} \end{equation}

\noindent Then, using Eq.~(\ref{relThetat}) in Eq.~(\ref{D1})
and Eq.~(\ref{area2}), our final expressions for the area and
angular-diameter distance are

\begin{equation} dA_{s}=\frac{\sin \Theta }{2l^{2}}\left|
\frac{\partial \Theta }{\partial \psi }\right| \left( 1-\sin
^{2}\!\psi \frac{l^{2}(1-2\sqrt{2}Ml)}{
l_{0}^{2}(1-2\sqrt{2}Ml_{0})}\right) ^{\frac{1}{2}}d\psi
d\gamma \label{area3} \end{equation}

\noindent and

\begin{equation} D_{A}^{2}=\frac{\sin \Theta }{2l^{2}\sin \psi
}\left| \frac{\partial \Theta }{\partial \psi }\right| \left(
1-\sin ^{2}\!\psi \frac{l^{2}(1-2\sqrt{2}Ml)}
{l_{0}^{2}(1-2\sqrt{2}Ml_{0})}\right) ^{\frac{1}{2}}.
\label{D2} \end{equation}

It should be noted that the only place where $D_{A}$ vanishes
is at $\sin \Theta =0$, namely, along the optical axis.  The
factor $\partial {\Theta }/ \partial \psi $ does not vanish
anywhere and diverges at $l=l_{p}$ at the same rate as the
factor $(1-\sin ^{2}\!\psi \frac{l^{2}(1-2\sqrt{2}Ml)}{
l_{0}^{2}(1-2\sqrt{2}Ml_{0})})^{1/2}$ approaches zero (See
Eq.~(\ref{Thetapsi}) in the appendix).

From Eq.~(\ref{J}) and Eq.~(\ref{D2}), we see that the square
of the angular-diameter distance is proportional to the
Jacobian of the lens mapping.  Because the angular diameter
distance appears in the denominator of the apparent
brightness, $S$, (see Eq.~(\ref{candle})), a point source
lying on the caustic will be infinitely magnified in  the
geometrical optics limit.  In addition, our expression for the
angular-diameter distance substituted into Eq.~(\ref{murel})
gives the relative magnifications for two lensed images.

If one observes two or more images in the directions $\{
\psi_i \}$, Eq.~(\ref{t}) can be used to define time of
arrivals, $\{t_i \}$.  The subtraction of two such times
defines a coordinate time delay, which can be converted into a
proper time delay along an observer's world line. 

Among other possible candidates to useful lensing observables,
which we have not concerned ourselves with, preliminary
calculations suggest that the distortion of the images of small
sources could be suitable for the application of the exact
formalism as developed in this particular section.

\subsection{Observables and the Parameter $l$}

In each of our calculations of observable quantities, the
non-measurable inverse radial parameter, $l$, plays an
essential role.  As mentioned earlier, this parameter should
be eliminated in terms of observable quantities, perhaps a
physical distance scale.

The most direct possibility is the angular-diameter distance
expression given by Eq.~(\ref{D2}).  We can see that it is not
a simple matter to invert the angular-diameter distance in
order to infer $l$ in terms of observables and lens
properties.  Nevertheless, Eq.~(\ref{D2}) is an implicit
relationship between $l$ and the observable $D_{A}$, and can
be solved numerically in local patches.

A second observational way to estimate the value of $l,$ is
via Eq.(\ref {murel}).  We have indicated that the ratio of
the brightness of a source in two images yields an implicit
equation for the source position $l$ through the distance
relationship, Eq.~(\ref{D2}). Hence, the parameter $l$ may be
replaced by $\mu_{12}$ in all calculations.

[An alternative approach, perhaps of only academic interest,
to the inverse radial distance $l$ can be obtained from the
redshift of the source in closed form.  If we assume that the
source and observer are at rest then $v_{s}^{a}=|g_{00}(x^{a}(
\alpha_1,\alpha _{2},x^{1}))|^{-1/2}(1,0,0,0)$ and
$v_{0}^{a}=|g_{00}(x_{0}^{a}|^{-1/2}(1,0,0,0)$, thus
$g_{ab}v^{a}\dot{X} ^{b}=|g_{00}|^{-1/2}g_{00}\dot{t}$ for
both $\omega _{s}$ and $\omega _{0}$.  Using the metric, we
also have $g_{00}\dot{t}=1/\sqrt{2}$ at both locations.  Thus

\begin{equation} \frac{\omega _{s}}{\omega _{0}}=\left(
\frac{1-2\sqrt{2}Ml_{0}}{1-2\sqrt{2}Ml }\right)
^{\frac{1}{2}}, \end{equation}

\noindent which is of course the standard gravitational
redshift~\cite{wald84} for Schwarzschild spacetime, expressed
in our notation.  Thus $l$ can be obtained as a function of
the ratio of the observed and source frequencies (and lens
mass $M$ and observers position $l_{0}$) by \begin{equation}
l=\frac{1}{2\sqrt{2}M}\left( 1-\frac{\omega _{s}^{2}}{\omega
_{0}^{2}}(1-2 \sqrt{2}Ml_{0})\right) .  ] \label{lobserved}
\end{equation}

\subsection{Image and Lens Properties}

If one is interested in learning properties of an unseen lens,
then the lens equation Eq.~(\ref{lenseq}) with Eq.~(\ref{D2}) can
be used in conjuction with knowledge of the image properties,
especially the brightness of the two main images.  Because the
brightness $S$ of the images is proportional to an inverse power
of the angular-diameter distance $ D_{A},$ (via
Eq.~(\ref{candle})), the brighter images will be observed when
the source is located near a caustic.

\begin{figure}
\centerline{\psfig{figure=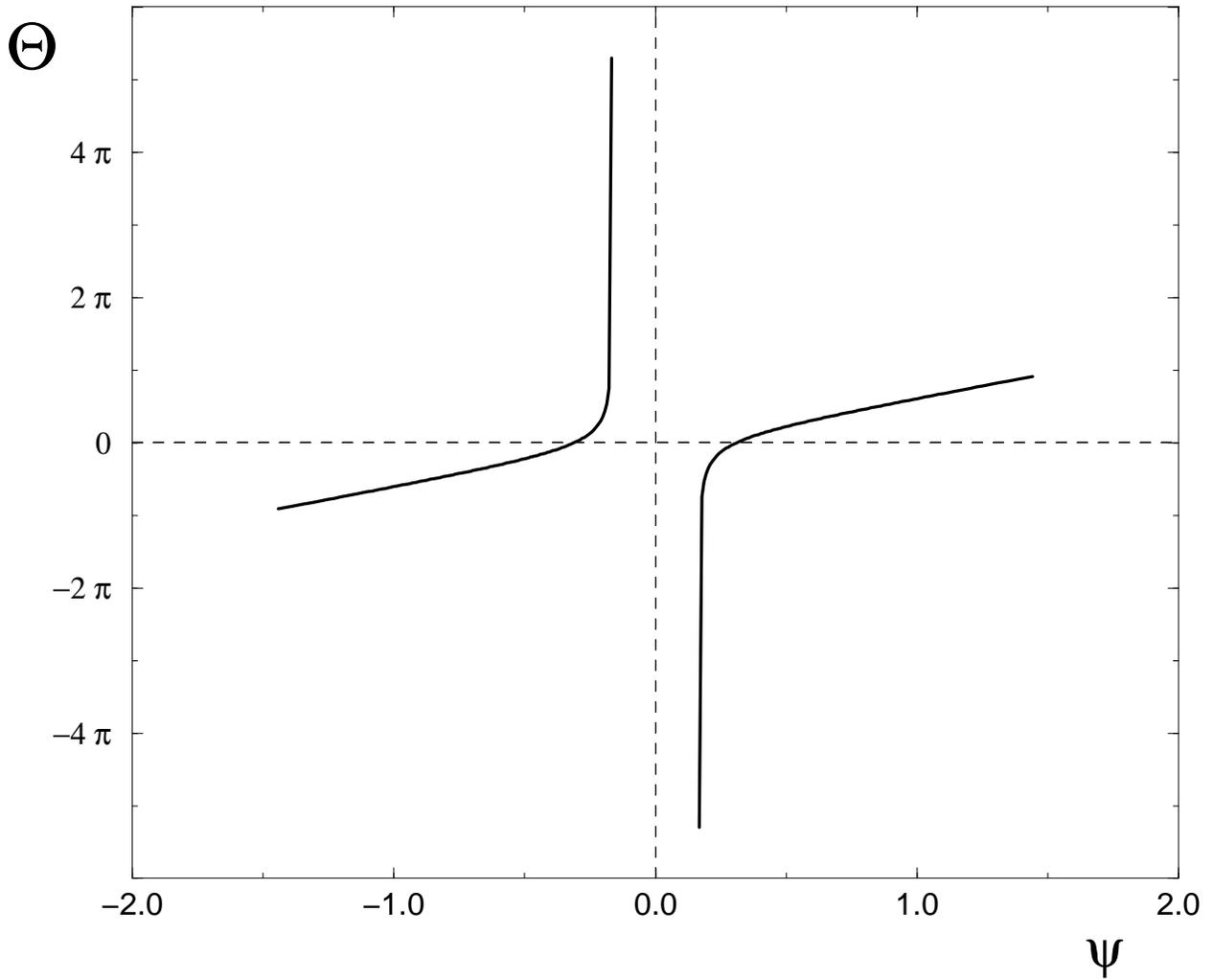,width=6.5in,angle=-90}}

\caption{The function $\Theta(l,l_0,\psi)$ at fixed values of
$l$ and $l_0$.  The angles on both axis are measured in
radians.  We have chosen $l=l_0=0.1/3\protect\sqrt{2}M$ with
$M=1$ for this figure.  This plot represents the exact lens
equation in the case that the $z-$axis is chosen as the
optical axis, defined by the observer and the lens.  }
\label{lenseqfigure} \end{figure}

We can see that the angular-diameter distance vanishes at
locations where either $M_{1}^{a}$ or $M_{2}^{a}$ vanish.  At
such locations, neighboring rays meet.  We have that
$M_{1}^{a}$ vanishes at $\Theta =0$, which means that
neighboring rays with the same value of $\psi $ meet along the
optical axis.  (In fact, all rays with that value of $\psi$
cross there, although only neighboring ones contribute to the
intensity of the image.)  This means that an observer in line
with the lens and a source would observe an extremely bright
perfect ring centered on the optical axis, at an angle $ \psi
$ that makes the function $\Theta $ vanish (the Einstein
ring).  From Fig.~\ref{lenseqfigure}, it can also be seen that
as the image angle approaches zero, the source angle,
$\Theta$, tends to infinity.  This means that there will be an
infinite number of Einstein rings, appearing in principle for
a black hole lens, at smaller and smaller angles -- one each
time that $\Theta $ passes through an integer number of turns,
i.e., $\Theta = 2\,n\,\pi$.  However, $\partial \Theta
/\partial \psi $ goes to infinity as well, which means that
the additional Einstein rings get much dimmer as the image
angle goes to zero.

If the source is not on the optical axis, two main images will
form, one on each side of the lens, because $\Theta $ is an
odd function of $\psi $.  One image will form at large
positive angle $\psi$, whereas the opposite image forms at
small negative angle $\psi $, and they will have different
brightness.  The images at smaller angles are dimmer, because
$ \Theta $ diverges steeply at the $3M$ radius, i.e.,
$\partial \Theta / \partial \psi$ is large.  As with the
Einstein rings, in addition to the two images there will be an
infinite number of other images, dimmer and at smaller angles
for a black hole lens.

On the other hand $M_{2}^{a}$ $=0$ would vanish at locations
such that $ \partial \Theta /\partial \psi =0$, where
neighboring rays with the same value of $\gamma $ meet.  These
lie on a plane containing the source, the observer and the
lens.  It can be seen from Fig.~\ref{lenseqfigure} that in the
case of a black hole, where the mass is contained within the
$3M$ radius, $\Theta $ is a monotonically increasing function
of $\psi $, and thus there are no points where $\partial
\Theta /\partial \psi $ vanishes.  Thus, we are not concerned
with these caustics.  These caustics do form in the case where
a spherical lens is modeled as a uniform dust sphere with
radius larger than $3M$, and lie on the lightrays that travel
through the mass, assuming the mass is transparent.

We can now use the lens equation Eq.~(\ref{lenseq}) to infer
properties of an unseen dark matter or blackhole lens.  Using
Eq.~(\ref{D2}) with Eq.~(\ref{lenseq}), and labeling the
optical axis as the $z-$axis, we have:

\begin{equation} \theta =\Theta (l(M,observables),l_{0},\psi )
\label{thetaobs}\end{equation}

\noindent where the $observables$ might be the brightness and
luminosity or (see below) the frequency ratios.  For example,
if an Einstein ring is observed at an angle $\psi _{1}$, then
we know both the image angle and the source's angular
location, i.e., $\theta =0$. Then Eq.~(\ref{thetaobs}) yields
a relationship between the mass $M$ and the inverse radial
location $l_{0}$ of the unseen blackhole.  This is not
particularly useful, but if another Einstein ring is observed
at an angle $\psi _{2}$, then we have two equations,

\begin{eqnarray} 0 &=&\Theta (l(M,observables),l_{0},\psi _{1}) \\ 0 &=&\Theta
(l(M,observables),l_{0},\psi _{2}) \end{eqnarray}

\noindent for the two unknowns $l_{0}$ and $M$ and we can infer
both the location and mass of the blackhole from lensing
observables.  Notice that this exact method is necessary in order
to treat multiple Einstein rings, since the standard weak-field
lens equation yields only one Einstein ring.  This fact has been
observed in~\cite{ellis}, and is emphasized in the following
section, where we compare the standard lens equation with the
exact approach. As for the applicability of the method, we refer
the reader to~\cite{ellis}, where magnifications and angular
locations of multiple Einstein rings have been calculated for the
case of the Milky Way's galactic blackhole.

\section{Comparisons of the exact and thin-lens equations}

\begin{figure}

\centerline{\psfig{figure=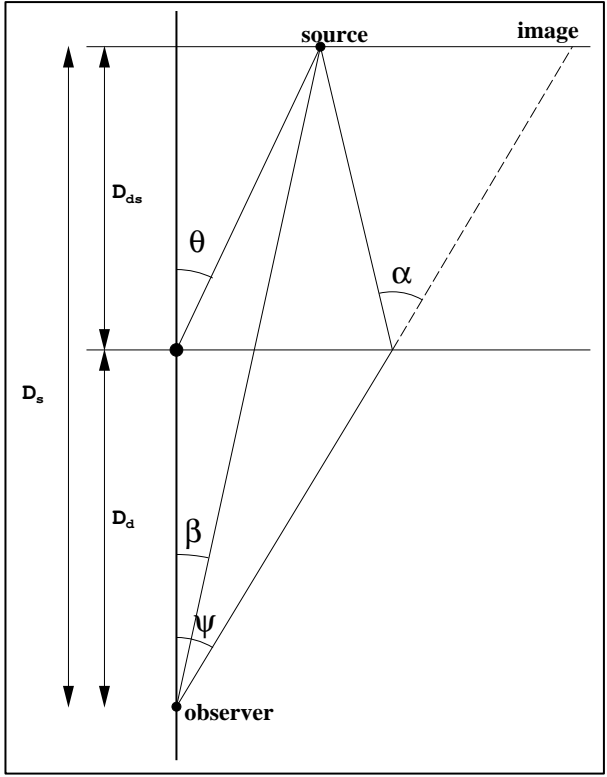,width=6.5in,angle=-0}}

\caption{The lens diagram for the case in which a background
is available.  The quantities
$\beta,\psi,\alpha,D_s,D_d,D_{ds}$ represent the Euclidean
angles and angular diameter distances in the background
space-time shown.  The quantity $\theta$ represents the
angular position of the source with respect to the $z-axis$ in
Schwarzschild coordinates if Schwarzschild spacetime is
thought of as superimposed on a flat background.  }
\label{lensdiagram} \end{figure}

The available approaches to lensing build on the view that the
lens is a perturbation on a given background.  The backgrounds
are normally taken as Minkowski spacetime or as a Friedmann
model.  For definiteness, here we restrict to a Minkowski
background.  Given the flat background, one can define the
locations of the source plane and the lens plane, at distances
$ D_s$ and $D_d$ from the observer, respectively, as shown in
Fig.~\ref {lensdiagram}.  One can further define the angular
location of the source on the lens plane $\beta$, and the
angular location of the image on the lens plane $\psi$.  One
can also define $D_{ds}$, the distance between the lens plane
and the source plane.  The thin-lens approximation consists in
considering the bending to take place at the lens plane, the
lightrays being otherwise straight lines.  The straight
lightrays bend through a bending angle $\alpha$.  The
thin-lens approximation is justified from the point of view
that the distances involved are much larger than the extent of
the gravitational field, and is normally accompanied by the
assumption that the image angles are small.

In our comparisons, we consider two different approximate lens
equations available.  One is the standard weak-field lens
equation~\cite{Ehlers}, and the other is a strong-field thin
lens equation obtained recently in~\cite{ellis}.  These two
lens equations differ essentially in the calculation of the
bending angle at the lens plane.

The standard weak-field approximation calculates the bending
angle via linearized Schwarzschild.  This results in small
bending angles, which justifies a further assumption that the
source angle $\beta$ is small.  Thus the standard weak-field
thin-lens equation is also a small-angle approximation, where
$\tan x=\sin x=x$ for $x=\beta,\psi,\alpha$.  The weak-field
thin-lens equation for a linearized Schwarzschild lens is

\begin{equation} \label{standard}
\beta=\psi-\frac{4MD_{ds}}{D_dD_s\psi} 
\end{equation} 

\noindent In our current notation, the angular location of
the source from the optical axis is $\theta$. We need to
transform $\beta$ into a function of $\theta$ in order to make
a comparison with the exact lens equation.  In this
approximation, however, since all angles are small, then from
Fig.~4.  $\beta/\theta=D_{ds}/D_s$, thus

\begin{equation} \label{weakbeta}
\beta=\frac{D_{ds}}{D_s}\theta.  \end{equation}

\noindent We also need to express the distances in terms of
the coordinates $(l, \theta)$. In this case, because of the
small-angle assumption, we have

\begin{eqnarray} D_d&=& \frac{1}{\sqrt{2} l_0}, \\ D_{ds} &=
&\frac{1}{\sqrt{2} l}, \\ D_s &= &\frac{1}{\sqrt{2} l_0}
+\frac{1}{\sqrt{2} l}.  \label{weakdist} \end{eqnarray}

\noindent Using Eq.~(\ref{weakbeta}) and Eq.~(\ref{weakdist})
with Eq.~(\ref{standard}), we obtain the weak-field thin-lens
equation in our current notation:

\begin{equation} \theta = \frac{l+l_0}{l}\left(\psi
-\frac{4\sqrt{2}M}{\psi} \frac{l_0^2}{ (l_0+l)}\right). 
\label{ourstandard} \end{equation}

\noindent This is to be compared with the lens equation,
Eq.~(\ref{theta}), in the case $ \theta_0=\pi$.  We can see
that $\theta=\Theta$ reduces to Eq.~(\ref{ourstandard}) in the
following manner.  We assume that the dimensionless quantities
$M~l \equiv \epsilon$ and $M~l_0 \le \epsilon$ are small and
make a Taylor series expansion of $\Theta \equiv \pi - \Delta$
in terms of $\epsilon$:

\begin{equation} \label{taylor} \Theta(l, l_\circ, \psi)
\approx \pi - \Delta(\epsilon = 0, b^\ast, l_\circ, l) +
\epsilon \left[\frac{\partial \Delta}{\partial \epsilon}
\right]_{\epsilon = 0} + \frac{\epsilon^2}{2} \left[ \frac{
\partial^2 \Delta }{\partial \epsilon^2} \right]_{\epsilon =
0} \equiv \pi - \Delta_\circ - \Delta_1 - \Delta_2.
\end{equation}

One can show that the zeroth and first order terms in the
Taylor series expansion, Eq.~(\ref{taylor}), yield
Eq.~(\ref{ourstandard}).  We can further derive the next term
in the expansion.  The second-order correction, $ \Delta_2 $,
is obtained to lowest order in $\psi$ by

\begin{equation} \label{delta2} \Delta_2 =
\frac{15~\pi~(M)^2}{4 D_d^2 \psi^2}, \end{equation}

\noindent which corrects Eq.~(\ref{ourstandard}) as

\begin{equation} \label{ourstandardsec} \theta =
\frac{l+l_0}{l}\left(\psi -\frac{4\sqrt{2}M}{\psi}
\frac{l_0^2}{ (l_0+l)} -\frac{15\pi M^2}{2\psi^2}
\frac{l_0^3}{l_0+l}\right).  \end{equation}

Equation~(\ref{ourstandardsec}) translates into a second-order
accurate version of the standard weak-field thin-lens equation
in the form

\begin{equation} \label{newapprox} \beta = \psi -
\frac{4M~D_{ds}}{D_d~D_s~\psi} - \frac{15\pi}{4} \frac{
M^2~D_{ds}}{D_s~D_d^2~\psi^2}.  \end{equation}

The strong-field thin-lens equation~\cite{ellis} calculates
the bending angle in exact Schwarzschild, for a lightray that
deviates a total angle $ \hat{\alpha}$ between the incoming
direction from infinity and the outgoing direction to
infinity.  The bending angle $\hat{\alpha}$ is not necessarily
small, and will increase to infinity as rays approach the $3M$
radius.  Thus there is no basis for a small angle
approximation of the angles $\beta$ and $ \hat{\alpha}$.  The
image angle $\psi$ is still considered small in the spirit of
the thin lens approximation, and the lens and source spheres
are still considered planes.  The strong-field thin-lens
equation is obtained in the same manner as the weak-field
thin-lens equation is, from trigonometry on the lens diagram,
but uses the exact trigonometric functions of $\beta$ and
$\hat{\alpha}$, rather than substitute the sines and tangents
by their angles.  The strong-field thin-lens equation is
Eq.~(1) in \cite{ellis}, namely

\begin{equation} \label{ellis} \tan\beta =
\tan\psi-\frac{D_{ds}}{D_s}\left[
\tan\psi+\tan(\hat{\alpha}-\psi) \right] \end{equation} with
\begin{equation} \label{hatalpha} \hat{\alpha} = 2\int_0^{l_p}
dl \frac{1}{\sqrt{ l_p^2(1-2\sqrt{2}Ml_p) -l^2
(1-2\sqrt{2}Ml)}} -\pi \end{equation}

\noindent Equation~(\ref{hatalpha}) is obtained from Eq.~(10)
in~\cite{ellis} by making the identifications $r\to
\frac{1}{\sqrt{2}l}$ and $r_0\to \frac{1}{\sqrt{2}l_p}$ .

Since we are interested in comparing with the exact lens
equations, we need to express $\beta, \hat\alpha$ and $D_d,
D_{ds}, D_s$ in terms of our coordinates.  In the first place,
since angles are not small then we have

\begin{equation} \label{ellisbeta}
\tan\beta=\frac{D_{ds}}{D_s}\tan\theta \end{equation}

\noindent and \begin{eqnarray} D_d&=& \frac{1}{\sqrt{2} l_0},
\\ D_{ds} &= &\frac{\cos\theta}{\sqrt{2} l}, \\ D_s &=
&\frac{1}{\sqrt{2} l_0} +\frac{\cos\theta}{\sqrt{2} l}. 
\label{ellisdist}\end{eqnarray}

\noindent Additionally, from Eq.~(\ref{Theta}), it is
straightforward to see that

\begin{equation} \label{ellisalpha} \hat{\alpha} = -\Theta( l,
l_0,\psi)|_{l=l_0=0}. \end{equation}

With Eqs.~(\ref{ellisbeta}), (\ref{ellisdist}) and
(\ref{ellisalpha}), Eq.~(\ref {ellis}) becomes

\begin{equation} \sin\theta
-\cos\theta\tan\big(\Theta|_{l=l_0=0}+\psi\big) -\frac{l}{l_0}
\tan\psi =0 \end{equation}

\noindent which can be manipulated to yield

\begin{equation} \label{ourellis} \theta = \psi
+\Theta|_{l=l_0=0} +\arcsin\left(\frac{l}{l_0}\tan\psi \cos
\big(\Theta|_{l=l_0=0}+\psi\big) \right). \end{equation}

\noindent Equation~(\ref{ourellis}) is the strong-field
thin-lens equation and is to be compared with the exact lens
equation $\theta=\Theta$.

\begin{figure}

\centerline{\psfig{figure=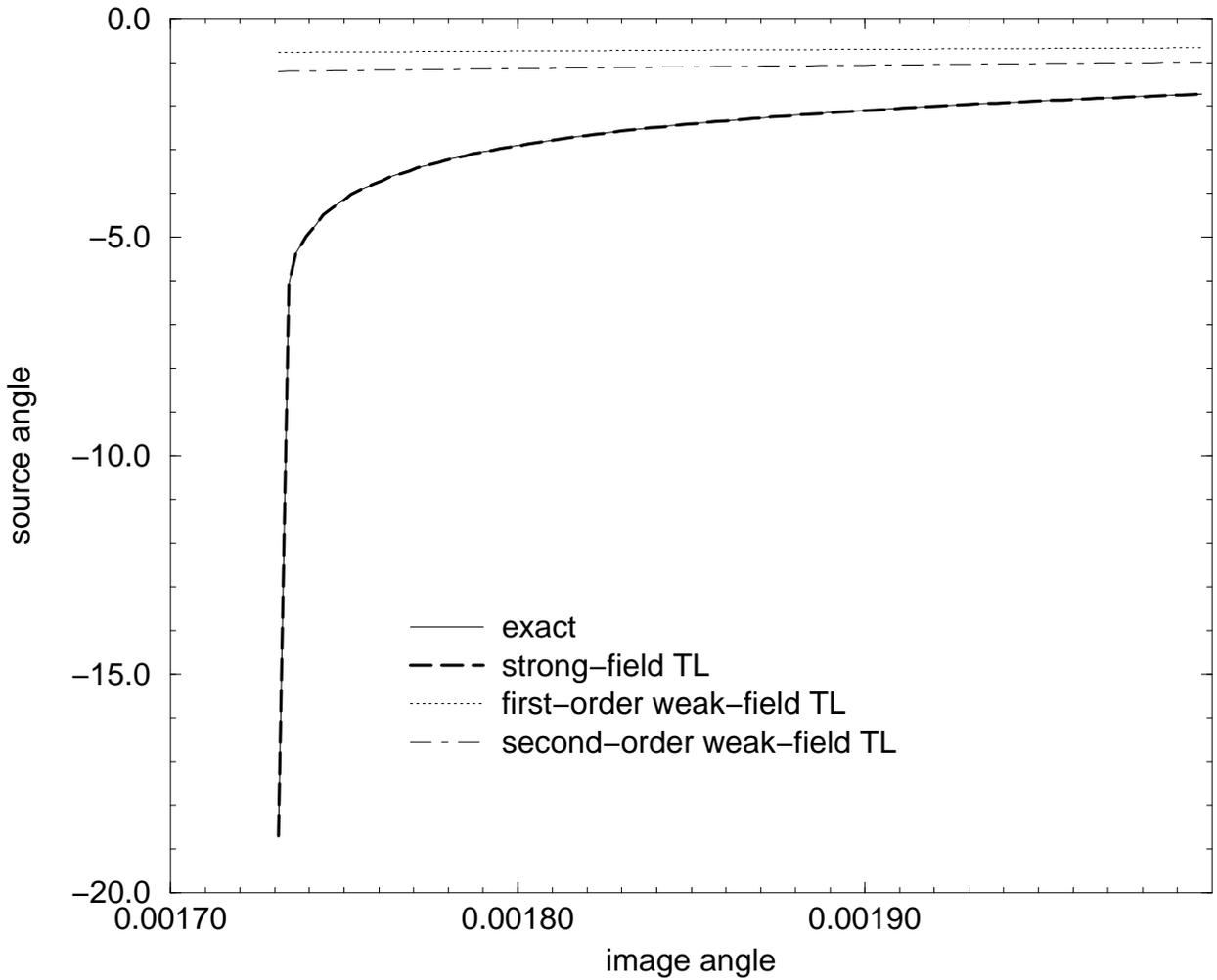,width=6.5in,angle=-90}}

\caption{A comparison of the available lens equations in the
literature.  The vertical axis represents the angular position
of the source in radians.  The horizontal axis represents the
observed angle $\psi$ \textbf{ in radians}.  We have taken the
observer and the source at the same distance from the lens,
$r=3000M$.  Here TL:=thin lens.  The standard
thin-lens-weak-field approximation is clearly inaccurate,
whereas the second-order correction to it does not do
significantly better.  However, the thin-lens-strong-field
approximation appears to be remarkably good, since at this
resolution it agrees with the exact lens equation up to one
part in a thousand.  } \label{comparison} \end{figure}

We can now plot our three lens equations and see how well they
compare.  We choose $l=l_0=\frac{0.001}{3\sqrt{2}M}$, which
corresponds to a distance of $ 3000M$ from the lens for both
the source and the observer and set $M=1$ for the plots.  We
look at small image angles, which corresponds to approaches
close to the $3M$ radius.  The plots are only made for
positive image angles.  Since the lens equations are odd
functions of $\psi$, the plot for negative $ \psi$ is a
reflection through the origin and can be inferred from
Fig.~\ref {comparison}.

We can see in Fig.~\ref{comparison} that the weak-field
equation (\ref{ourstandard}) deviates significantly from the
exact equation both at first-order accuracy and at
second-order accuracy, Eq.~(\ref{ourstandardsec}), as
expected.  The main reason for the second-order correction not
to behave significantly better than the first-order accurate
equation is that none of the perturbative terms blow up at the
$3M$ radius, whereas the exact equation does.  The
perturbative approach is clearly inappropriate for the
strong-field regime of a black hole.

However, the strong-field thin-lens equation
Eq.~(\ref{ourellis}) agrees remarkably well with the exact
equation.  In our example, we have an error of less than one
part in a thousand, for source angles as large as about six
turns around the lens.  Notice that 3000M corresponds to only
$4500$km for a stellar black hole, nowhere near the galactic
distances expected for observed lensing.  The thin-lens
approximation appears to do very well at these relatively
small distances and will almost certainly do much better at
galactic and extragalactic distances.

One reason why Eq.~(\ref{ellis}) performs so accurately is the
tremendous distances involved.  The other reason is, however,
that the exact bending angle through infinite distances is
used.  Most of the effort in using Eq.~(\ref{ellis}) is
involved in evaluating the integral expression of the bending
angle and its derivatives, which involves the same amount of
work required to do the exact lens equation, as shown by our
comparison above, Eq.~(\ref{ourellis}).  Thus the efficiency
in terms of computational cost is not very high. Considering
that the price is to break with the covariance and
non-linearity of general relativity, one might even find it
justified to give up the thin-lens approximation in favor of
the exact lens equation in the case of strong, spherically
symmetric gravitational fields.

A valuable, second comparison is the time delays predicted by the exact and thin lens
approaches.  Time delays are now a very important tool in astrophysics and provide a means
to estimate the Hubble constant independent of previous methods.  We will define a time
delay as the observer's elapsed proper time between the arrival of a signal along two
distinct paths connecting the source and observer.

When the weak field, thin lens approach is applied to the
Schwarzschild lens, generically, two paths connect any source
and observer, one passing on each side of the lens.  This
results from the lens equation, Eq.~(\ref{thetaphi}),
admitting two values for $\psi$ for given values of $\beta$,
$M$, and distance parameters.  The only exception is when the
source, lens, and observer lie along the same radial line.  In
this case an Einstein ring is formed, and the time delay is
zero by symmetry.  The time along each of the two paths is
found by integrating

\begin{equation}t^{tl} = \int \, dl \, 
\left(1 + \frac{2M}{r(l)}\right) \end{equation}

\noindent along the trajectory determined by the weak field
thin lens equation.  In the integration, $l$ is the Euclidean
length along the thin lens trajectory, and $r(l)$ represents
the Euclidean distance from the origin to a point along the
path.  The thin lens time delay, $\Delta t^{tl}$, is the
difference between the two times.

In the exact approach, there are an infinite number of pairs
of geodesics connecting the source and observer because
geodesics may wrap around the black hole many times.  However,
there are only two geodesics which do not circle the lens;
these geodesics, which can be distinguished from the others,
are the analogue of the thin lens paths.  The exact time
delay, $\Delta t^e$, is numerically computed from
Eq.~(\ref{t}) by taking the difference between the times
elapsed along these two geodesics.

\begin{figure}

\centerline{\psfig{figure=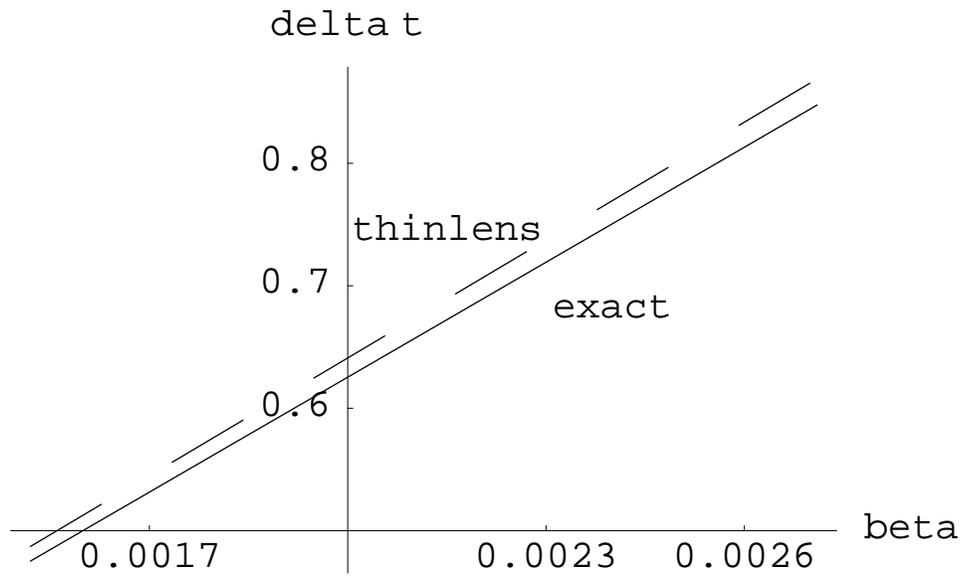,width=5in,angle=-0}}

\caption{A plot of the exact and thin lens time delays against
the source angle $\beta$ in radians.  Here, the source and
observer are roughly $3000 M$ from the lens.}  \label{timefig}

\end{figure}

Figure \ref{timefig} shows the exact and weak field thin lens
time delays given the same observer and source location.  The
time delays are given in geometrical units, and the
$\beta$-axis is given in radians.  In this calculation, the
observer is located at $l_0 =\frac{0.001}{3\sqrt{2}M}$, or a
distance of $3000 M$, and $\theta = 0$.  Sources are located
in a fixed source plane whose distance from the lens along the
optical axis is also $3000M$, as in the standard thin lens
picture.  The time delays are plotted against $\beta$, the
source angle used in the thin lens equation.

At $\beta = 0$, Einstein rings are formed, and each time delay
will be zero.  As $\beta$ increases, the thin lens time delay
slightly overestimates the true value, as seen in the figure. 
This represents the general behavior for a single
Schwarzschild lens at larger distances, but the effect remains
fairly small.  For a single Schwarzschild lens with a mass of
$2.5 \times 10^{12}$ solar masses, at redshift $z = 0.5$ with
a source at $z = 1.0$ and $\beta = 0.5''$, the overall exact
time delay is close to $400$ days, while the error introduced
by using the thin lens approximation is less than an hour.

The time delays computed in the thin lens approximation remain
quite accurate in our comparisons because the light rays do
not feel a strong gravitational field along their
trajectories.  This is why the thin lens fails so dramatically
in the observation angle calculation presented in
Fig.~\ref{comparison}, while fairing well in the time delay
comparisons.

Despite these results, it may be incorrect to assume that the
thin lens method of computing time delays remains accurate for
more complex lensing situations.  The first of two possible
problems is that the weak field assumption may not always be
valid along an observed null geodesic.  It may be very
difficult to identify lens candidates whose null geodesics
have undergone interactions with strong gravitational fields,
however, if such candidates are found, the thin lens
methodology will most likely prove quite inaccurate. 
Secondly, it is not known how accurate the time delay
computations will be in more complex lensing scenarios,
especially when the geodesics are bent in multiple lens
planes.

\section{Discussion}

To date, virtually all applications of gravitational lensing
have utilized the thin lens approximation.  Although the thin
lens method has proven a quick and useful tool, it can not be
applied to high curvature regions.  The failure of the next
order correction to the thin lens equation,
Eq.~(\ref{newapprox}), as contrasted with the ability of the 
strong field thin lens equation~\cite{ellis}, to capture the
divergence of the observation  angle $\Theta$ in
Fig.~\ref{comparison} emphasizes the point that some combination
of exact  methods are required in such situations.

We are putting forward the idea that exact gravitational
lensing may not be just a purely academic exercise.  In fact,
we see that virtually all of the observationally relevant
quantities can be determined analytically in the exact method
with relative ease for a  Schwarzschild spacetime.  In this
paper, we found straight-forward expressions for the time
delays, observation angles, angular-diameter distances, and
relative magnifications.  These analytic expressions can be
used in  comparison calculations or model building with great
computational prowess, power, or time.

The limited testing of the thin-lens methodology in this paper
has indicated that there are regions (possibly not yet
observable) where the thin-lens method fails in a highly 
symmetric case. In ongoing calculations, we are exploring the
accuracy of the thin lens  approximation under a broad range
of conditions, including those in which there may be  stronger
fields or multiple lenses present. These calculations will
require a combination  of analytic and numerical results. A
precedent for this kind of work was set by Rauch and 
Blandford~\cite{Blandford}, who studied the null geodesic
equations (or the exact lens  equations in our terminology)
for Kerr spacetime and found the caustics of the lightcone. 
Our sense from these comparisons is that the error in the thin
lens method for the time  delays and observation angles will,
in some cases, be appreciable~\cite{KNP}.

We are also interested in an  issue regarding the comparison
of the sizes of two different  types of corrections to the
thin lens equations. On one hand, within the framework of the 
thin lens methodology, there are corrections due to the
structure of the mass distribution  (of the lens) over that of
the monopole moment. On the other hand, even for the monopole 
case, there are differences between the predictions of the
exact lens equations and the  thin lens equations. The issue
is whether the sizes of these corrections are comparable.

\acknowledgments

This research has been supported by the NSF under grants No. 
PHY-9803301, PHY-9722049 and PHY-9205109.

\appendix

\section{}

An alternative derivation of Eq.~(\ref{relThetat}), by direct
computation, is obtained by taking the derivative of $\Theta $
and of $t$ with respect to $ \psi $.  Because both $t$ and
$\Theta $ depend on $\psi $ only through the point of closest
approach $l_{p}$, we actually need to calculate $\partial t/
\partial l_{p}$ and $\partial \Theta /\partial l_p$.  We
have 

\begin{eqnarray} 
	\frac{\partial \Theta }{\partial l_{p}}
   &=&
	+\int_l^{l_0} 
	 \frac{l_p(1-3\sqrt{2}Ml_p)dl'}
	      {\Big( l_p^2(1-2\sqrt{2}Ml_p)
		    -l'^2(1-2\sqrt{2}Ml')\Big)^{3/2}} \nonumber \\
   & &
	+\lim_{\epsilon \to 0}
	\Bigg\{ 
	       2\int_{l_0}^{l_p -\epsilon}  
		\frac{l_p(1-3\sqrt{2}Ml_p)dl'}
		     {\Big(l_p^2(1-2\sqrt{2}Ml_p)
			  -l'^2(1-2\sqrt{2}Ml')\Big)^{3/2}}\nonumber \\
   & &\hspace{1.5cm}
	-\left.\frac{2} 
		    {\sqrt{l_p^2(1-2\sqrt{2}Ml_p)
			  -l'^2(1-2\sqrt{2}Ml')}} 
	\right|_{l'=l_p-\epsilon }
	\Bigg\} 				 \label{Thetapsi}  
\end{eqnarray} 

On the other hand we
also have 

\begin{eqnarray} 
    	\frac{\partial t}{\partial l_p}
   &=&
	-\frac{1}{\sqrt{2}}\int_l^{l_0}
	\frac{l_p(1-3\sqrt{2}Ml_p)}
	     {\sqrt{l_p^2}(1-2\sqrt{2}Ml_p)}
	\Big(l_p^2(1-2\sqrt{2}Ml_p)
	     -l'^2(1-2\sqrt{2}Ml') \Big)^{3/2}dl' \nonumber \\
   & &  
	-\lim_{\epsilon \to 0}\Bigg\{
	\frac{2}{\sqrt{2}}\int_{l_0}^{l_p- \epsilon}
	\frac{l_p(1-3\sqrt{2}Ml_p)}
	     {\sqrt{l_p^2(1-2\sqrt{2}Ml_p)}
	\Big(l_p^2(1-2\sqrt{2}Ml_p)
	     -l'^2(1-2\sqrt{2}Ml')\Big)^{3/2}} dl'\nonumber \\ 
   & &\hspace{1.5cm}+\left.
	\frac{2}{\sqrt{2}l'^2(1-2\sqrt{2}Ml'))}
	\sqrt{\frac{l_p^2(1-2\sqrt{2}Ml_p)}
	     {l_p^2(1-2\sqrt{2}Ml_p)
	     -l'^2(1-2\sqrt{2} Ml')}}
	\right|_{l'=l_p-\epsilon }\Bigg\} \label{pretpsi} 
\end{eqnarray} 

\noindent Clearly the last term inside the braces under the
limit sign in (\ref {pretpsi}) can be substituted with

\begin{equation} \left. 
\frac{2}{\sqrt{2l_{p}^{2}(1-2\sqrt{2}Ml_{p}))}}\frac{1}{\sqrt{
l_{p}^{2}(1-2\sqrt{2}Ml_{p})-l'^{2}(1-2\sqrt{2}Ml')}}\right|
_{l'=l_{p}-\epsilon }  \end{equation} 

\noindent since the difference vanishes in the limit $\epsilon
\to 0$.  We can thus see that $\partial t/\partial l_{p}$ and
$\partial \Theta /\partial l_{p}$ are proportional to each
other via 

\begin{equation} \frac{\partial t}{\partial
l_{p}}=-\frac{1}{\sqrt{2l_{p}^{2}(1-2\sqrt{2}
Ml_{p})}}\frac{\partial \Theta }{\partial l_{p}}. 
\label{relThetatlp} 
\end{equation} 

\noindent This implies Eq.(\ref{relThetat}).

\newpage

\newpage

\newpage

\newpage  

\newpage

\newpage

\end{document}